# Load & Backhaul Aware Decoupled Downlink/Uplink Access in 5G Systems

Hisham Elshaer[*,#], Federico Boccardi[*], Mischa Dohler[#] and Ralf Irmer[*]
[*]Vodafone Group R&D. Newbury, UK
[#]King's College London. London, UK
Email: {hisham.elshaer, federico.boccardi, ralf.irmer}@vodafone.com
mischa.dohler@kcl.ac.uk

*Abstract* — Until the 4[th] Generation (4G) cellular 3GPP systems, a user equipment's (UE) cell association has been based on the downlink received power from the strongest base station. Recent work has shown that – with an increasing degree of heterogeneity in emerging 5G systems – such an approach is dramatically suboptimal, advocating for an independent association of the downlink and uplink where the downlink is served by the macro cell and the uplink by the nearest small cell. In this paper, we advance prior art by explicitly considering the cell-load as well as the available backhaul capacity during the association process. We introduce a novel association algorithm and prove its superiority w.r.t. prior art by means of simulations that are based on Vodafone's small cell trial network and employing a high resolution pathloss prediction and realistic user distributions. We also study the effect that different power control settings have on the performance of our algorithm.

## I. INTRODUCTION

Driven by an increasing density of small cells in heterogeneous 4G systems, it was recently shown that the traditional strategy of handing over up (UL) and downlink (DL) simultaneously based on downlink received power is significantly capacity suboptimal. Indeed, the UL performance gains for cell-edge users due to decoupling the DL and UL cell association were consistently shown to be in the order of 200-300% [4].

These capacity improvements in the UL are very timely since the UL traffic has been growing over past years with an unprecedented rate. This trend is driven by new applications which generate symmetric traffic, such as real-time gaming and video calls. In addition, the emerging array of social networking applications as well as machine-to-machine technologies generates more UL traffic than DL in an uncorrelated fashion. The optimization of the UL, particularly for disadvantaged users at the cell edge, is thus of highest important to a consistent quality of experience in emerging 5G systems.

The decoupling is facilitated by the fact that the degree of heterogeneity has increased dramatically over past years and is expected to grow further as part of 4G and 5G rollouts. This shift from a single-tier homogeneous network towards multi-tier heterogeneous networks (HetNets) composed of different types of small cells (Micro, Pico and Femto) comes along with the unique opportunity to have ample connections available at any point and time. This, in turn, facilitates our purpose of decoupling the UL from the DL and the thereby achieved performance gains.

The concept is shown in *Figure 1* where the Small Cell (Scell) has DL and UL cell borders which are defined by the DL received power and pathloss respectively; a UE between these two borders will tend to connect to the Scell in the UL and to the Macro cell (Mcell) in the DL as shown in the figure.

Some prior art is emerging in this field [4, 5], but have so far assumed that the cell association strategy is based on the link quality in each direction. That is, the decision to handover the DL is (and has been) based solely on the DL received power; whereas the decision to handover the UL is based solely on the UL pathloss. The system assumptions were to some extent ideal in that neither the cell load nor the backhauling capabilities have been taken into account – both of which have an impact onto the actual performance gains under more realistic operating conditions. This shortcoming is addressed in this paper at hand, where we proceed to outline prior related art as well as a summary of our technical contributions.

### A. Related Work

The concept of downlink/uplink decoupling (DUDe) has been discussed as a major component in future cellular networks in [1]-[3]. In [3], in particular, DUDe is considered as a part of a broader "device-centric" architectural vision, where the set of network nodes providing connectivity to a given device and the functions of these nodes in a particular communication session are tailored to that specific device and session. In [6] and [7], backhaul aware cell association was considered but only from a DL perspective. In [8] and [9], load aware cell association was studied in the DL as well. In [10], the authors study UL cell association in a game theoretic approach to optimize the packet success rate of the UEs.

### B. Contributions

In contrast to prior art, this paper focuses on the cell association algorithm where we argue that UL pathloss alone is not sufficient to efficiently apply DUDe. Notably, the association algorithm ought also to consider the overall load of the cell(s). Furthermore, since DUDe requires significant backhauling support, we also condition association with backhauling capacity. Therefore, instead of taking the decision based only on link quality, the system now considers the link quality, the cell load and the cell backhaul capacity. We then use a realistic scenario of a cellular network based on Vodafone's real-world planning/optimisation tools which, we believe, adds a lot of value and credibility to this work. We give special attention to UL power control where we show that the performance depends greatly on the power control settings. We use a flow level traffic model that is more realistic than the full buffer model considered in prior art. The results are then discussed and evaluated in great details, thus offering unique insights into the performance trends of the emerging decoupling concept.

The remainder of the paper is organized as follows. In Section II, the system model is briefly introduced. In Section III, we describe the cell association algorithm which is extending prior art. Section IV introduces the simulation setup and results are discussed in great details in Section V. Finally, the conclusions are drawn in Section VI.

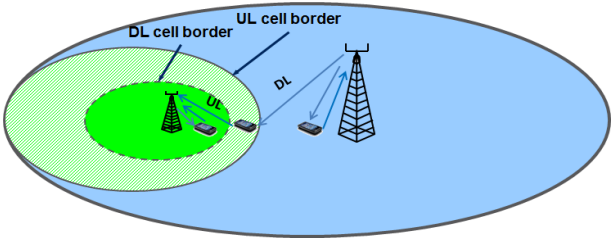

Figure 1. Illustration of the DUDe concept.

## II. SYSTEM MODEL

We consider the UL of a HetNet where, as deployment setup, we use the Vodafone LTE small cell test bed network that is up and running in the London area. The test network covers an area of approximately one square kilometre. We use this existing test bed to simulate a relatively dense HetNet scenario.

The considered network is shown in *Figure 2* where the black shapes are Macro sites and the red circles are Small cells where in total we have B cells.

We consider a realistic user distribution based on traffic data from the field trial network in peak times where the total number of users is $N_u$.

Network traffic is modelled on a flow level where flows represent individual file or data transfers e.g. video, audio or generic file uploads. This model reflects a much more realistic traffic model than the full buffer model considered in [4]. We assume that a flow of size $\rho$ arrives to a UE's queue after a certain period "wait time" $T_W$. $T_W$ and $\rho$ are exponentially distributed with certain mean values. UEs experience a different $T_W$ each time a flow transmission is finished.

The radio link quality is determined by many factors including pathloss, fading, interference, and transmit power of the UEs. The UL SINR of UE $i$ connected to BS $j$ is given by:

$$SINR_{ij} = \frac{h_{ij} P_i}{N+I}, \tag{1}$$

where $P_i$ is the i[th] UE transmit power, $h_{ij}$ incorporates pathloss, shadowing and fast fading between UE i and BS j, $N$ is the noise power and $I$ is the UL intercell interference. We characterize the achievable data rate using the Shannon formula where $BW$ represents the system bandwidth:

$$C_{ij}^{Access} = \text{BW} \log_2(1 + SINR_{ij}). \tag{2}$$

Uplink power control for the UEs follows the 3GPP specifications [11], where we consider open loop power control which is given by:

$$P_{UE} = min\{P_{MAX},\ 10\log_{10}(M) + P_0 + \alpha.L\ \}, \tag{3}$$

where $P_{MAX}$ is the maximum permittable transmit power of the UE, $M$ is the number of physical resource blocks (PRBs) assigned to the UE, $P_0$ is a normalized power, $\alpha$ is the pathloss compensation factor and $L$ is the pathloss towards the serving cell.

However, the power control algorithm does not account for inter-cell interference which, as we will show in the results, affects greatly the UL performance. The effect is more pronounced when load balancing takes place since UEs connect to a suboptimal cell so they are more vulnerable to interference. Therefore we will use an interference aware power control algorithm which sets a limit to the transmit power of the UEs depending on the interference level that the UE causes to the closest neighboring cell. Similar algorithms have been proposed in the literature such as [13]. The algorithm is as follows:

$$P_{UE} = \min\{P_{MAX}, 10\log_{10}(M) + P_0 + \alpha.L, I_0 + L_s + 10\log_{10}(M)\}, \tag{4}$$

where $I_0$ represents the UL interference power spectral density (PSD) target for the UE and $L_s$ is the pathloss towards the most interfered cell by the UE. This allows us to control the interference level in the system by changing $I_0$.

In a real world deployment, the Scell backhaul is always an issue since outdoor Scells are usually mounted on street furniture where there is no guaranteed wired connection or line-of-sight to the Mcell. Furthermore with the increasing bit rates provided by access technologies the bottleneck is moving slowly from the access network to the backhaul. We consider that all cells in the test network have a limited backhaul capacity $C_j^{bk}$ where, naturally, Scells would have tighter backhaul constraint than Mcells.

## III. CELL ASSOCIATION ALGORITHM

In our previous study [4] we have considered the UL cell association to be based on pathloss (PL) which showed very high performance improvements that were mainly due to the load balancing effect and the improved link quality of the UEs. We extend this approach to include the cell load and backhaul capacity in the decision criterion; consequently instead of taking the decision based only on link quality the UE considers the link quality, the cell load and the cell backhaul capacity. This approach makes sense since in real networks users are distributed in a non-uniform way where a UE that is close to a congested cell might be better off connecting to a cell that is further but less congested.

We consider a cell association criterion that was considered in [8] in the DL. We extend this by applying it to the UL and including the backhaul capacity so that the optimal BS chosen by UE $i$ is given by $s(i)$:

$$s(i) = arg\ max_{j \in B}\ (1 - \eta_j)\ C_{ij}^{Max}, \quad (5)$$

where $C_{ij}^{Max} = \min\{C_{ij}^{Access}, C_j^{bk}\}$, $\eta_j$ is the $j^{th}$ BS load which is reflected in [8] as being the average resource utilization per cell. We found that this approach for $\eta_j$ works fine in the DL whereas in the UL the situation is different since the UEs are power limited which means that a UE with bad channel conditions would not be able to transmit on a large number of resource blocks. This would result in a low utilization of the resources of the cell even though this cell could be serving many UEs. Therefore the cell utilization is a poor metric to characterize the cell load in the UL and we resort to a different way of estimating the load. Notably, since the flow arrival is exponentially distributed and assuming the system to be stationary, the stationary distribution of the number of flows $N_j$ is identical to that of an M/GI/1 multi-class processor sharing system [12]. The average number of flows is then given by $[N_j] = \frac{\eta_j}{1 - \eta_j}$, which yields $\eta_j = \frac{E[N_j]}{E[N_j] + 1}$. Inserting $\eta_j$ into (5) yields:

$$s(i) = arg\max_{j \in B}\ \frac{C_{ij}^{Max}}{E[N_j] + 1}. \quad (6)$$

The cell association criterion in (6) will be used for the rest of the paper. We consider a fully distributed algorithm where the main idea is that a UE does not need to stay connected to one BS in the UL all the time. Instead a UE can keep its anchor DL cell and every time the UE has data (flow) to transmit in the UL, the UE connects to the cell with the highest criterion according to (6).

The algorithm thus functions as follows: The BSs broadcast their load $E[N_j]$ and backhaul capacity $C_j^{bk}$. All UEs in the system start with an exponentially distributed wait time ($T_w$) after which a UE has a flow of size $\rho$ to transmit. The UE uses the criterion in (6) to find the best cell to connect to and after finishing its transmission the UE disconnects from the cell and goes idle for another random period $T_w$; thereupon the operation is repeated. The steps are detailed in Algorithm 1.

---

**Algorithm 1:** Load/backhaul aware UL cell association

1. BSs broadcast $E[N_j]$ and $C_{bk}$ periodically.
2. UEs (1… $N_u$) are idle for a random $T_w$(1… $N_u$).
3. for Number of subframes
4.   for each idle UE
5.     if $T_w$ = 0
6.       UE_queue = ρ.
7.       UE connects to BS (i) according to (6)
8.       UE is scheduled in BS (i) until UE_queue = 0.
9.       UE goes idle for a random $T_w$.
10.     else
11.       $T_w$ = $T_w$ - 1
12.   end for
13. end for

---

## IV. SIMULATION SETUP

In our simulations we use the deployment setup of the Vodafone LTE test network in the London area. The setup consists of 5 Mcells and 21 outdoor Scells as shown in *Figure 2*. Our propagation model is based on a high resolution 3D ray tracing pathloss prediction model. The model takes into account clutter, terrain and building data. This guarantees a realistic and accurate propagation model. The user distribution is based on traffic data extracted from the real network.

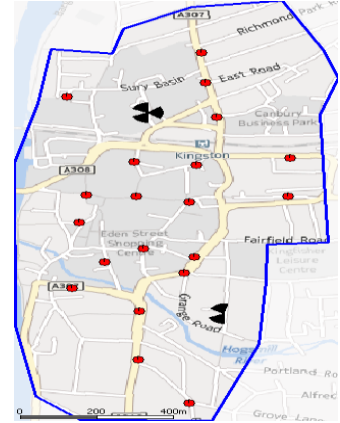

Figure 2. Vodafone small cell LTE test network in London.

We consider three power control settings:
- Loose power control with full pathloss compensation. We use (3) where $(\alpha, P_0)$ are set to (1, -80 *dBm*). This is referred to as *Setting 1*.
- Conservative power control with partial pathloss compensation. We use (3) where $(\alpha, P_0)$ are set to (0.6, -70 *dBm*). This is referred to as *Setting 2*.
- Interference aware power control where we use (4) and set $(\alpha, P_0)$ to (1, -80 *dBm*) and $I_0$ to -100 *dBm*.

Table 1. Simulation Parameters

| Operating frequency | 2.6 GHz (co-channel deployment) |
|---|---|
| Bandwidth | 20 MHz (100 frequency blocks) |
| Network deployment | 5 Mcells and 21 Scells distributed in the test area as shown in Figure 2. |
| User distribution | 330 UEs distributed according to traffic maps read from a live network |
| Scheduler | Proportional fair |
| Simulation time | 10 seconds (10,000 subframes) |
| Traffic model | Flow level traffic. Mean flow size = 1 Mbit. Mean wait time = 100 ms. |
| Propagation model | 3D ray-tracing model |
| Max. transmit power | Mcell = 46 dBm, Scell power = 30dBm, UE = 20 dBm. |
| Antenna system | Macro: 2Tx, 2Rx, 17.8 dBi gain Pico: 2Tx, 2Rx, 4 dBi gain UE: 1Tx, 1Rx, 0 dBi gain |
| UEs mobility | Pedestrian (3km/h) |
| Supported UL modulation schemes | QPSK, 16 QAM, 64 QAM |

We compare 3 UL cell association cases:
- Cell association based on the DL Reference Signal Received Power (RSRP) which is the conventional LTE procedure [11]. This case is termed **DL-RSRP**.
- Cell association based on the pathloss which represents the DUDe algorithm as considered in [4] and is termed as **DUDe**.
- Cell association based on Algorithm 1 which considers the cell load and backhaul capacity on top of the conventional DUDe. This case is termed **DUDe-Load**.

As pointed out before, all the results in the next section will focus on the UL performance. The simulation parameters are listed in Table 1 where we consider an outdoor LTE deployment.

## V. RESULTS & DISCUSSIONS

Initially we assume having an ideal backhaul (i.e. no limit on the backhaul capacity) on all the cells in order to study the load balancing effect. We start by comparing the throughput results with different power control settings according to *Setting 1* and *Setting 2*.

The throughput results for the three cases in comparison are shown in *Figure 3*. Comparing DUDe to DL-RSRP, we see similar gains as in [4] where the $5^{th}$ and $50^{th}$ percentiles are increased by more than 100% and 150% respectively for both power settings. The gains are due to the load balancing effect of DUDe and the better link quality as UEs connect to the cells to which they have the lowest PL. The $90^{th}$ percentile throughput is less in DUDe than DL-RSRP as in the latter case only a few UEs are served by the Scells; therefore these UEs achieve a high throughput.

We notice also that DUDe-Load is more affected, in terms of $5^{th}$ and $50^{th}$ percentiles, by the change in the power settings than DUDe. This is due mainly to the fact that UEs connect to suboptimal cell in terms of PL due to the load balancing effect which makes these UEs more vulnerable to interference and more affected by the other UEs transmit power.

We then compare DUDe and DUDe-Load starting by *Setting 1* where we see that the $5^{th}$ percentile throughput is reduced by about 20% in the DUDe-Load case while the $50^{th}$ percentile is increased by 40% compared to DUDe. The loose power control causes the interference level to increase which has a negative effect on the cell edge UEs as explained below.

This result shows that cell edge UEs ($5^{th}$ percentile) are better connected to a loaded cell to which they have the better link quality than connecting to an unloaded cell with a worse channel. On the other hand the $50^{th}$ percentile UEs can afford a reduced channel quality and with the higher power headroom they actually achieve a high gain by using the extra resources provided by the load balancing effect of DUDe-Load. Finally, the figure also shows a loss of about 20% in the $90^{th}$ percentile throughput which is logical since load balancing is always a trade-off between peak and (cell-edge/average) throughput.

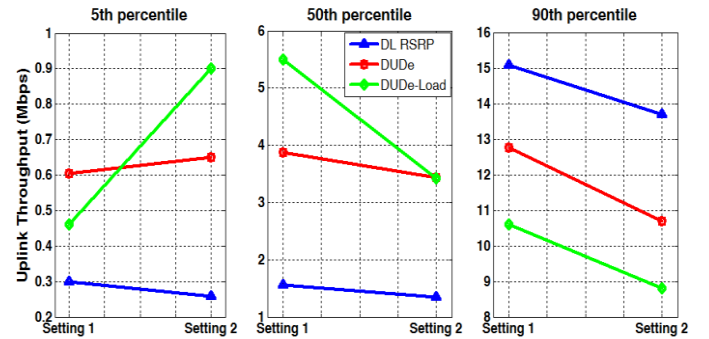

Figure 3. Throughput percentiles for the three cases with power control Setting 1 and 2.

Then we compare DUDe and DUDe-Load for *Setting 2* where the $5^{th}$ percentile throughput in DUDe-Load is improved by about 40% over DUDe whereas the $50^{th}$ percentile throughput is almost the same. This result shows how power control affects the network performance greatly. The used power control scheme sets a lower limit on the transmit power of the UEs than the one used in *Setting 1*; this causes the UL interference level in the network to be lower than the previous case which, in turn, allows the cell edge UEs to achieve a higher throughput when connected to a suboptimal cell in terms of pathloss.

On the other hand, the $50^{th}$ percentile UEs do not achieve a higher throughput with the load balancing effect due to the lower bound on the UEs transmit power. These UEs hence might not be able to use all the resources available to them; therefore, these UEs achieve a relatively low gain from the higher resource availability whereas the lower link quality to the suboptimal cell reduces the throughput. Consequently, both effects almost even out and there is no gain in terms of 50th percentile throughput.

The main message in *Figure 3* is that cell edge UEs are mostly interference limited whereas 50th percentile UEs are power limited so having power control *Setting 1* would benefit the 50th percentile UEs but would be harmful for cell edge UEs while power control *Setting 2* has the opposite effect.

*Figure 4* shows a CDF of the variance of the UEs UL SINR over time for *Setting 1* where interference is quite high. DUDe shows an average reduction of variance of about 10dB compared to DL-RSRP whereas DUDe-Load shows an even lower average variance of about 15dB compared to DL-RSRP. The lower variance reflects a more stable interference scenario in DUDe where the lower variance of DUDe-Load results from the improved load balancing effect which improves the resource utilization and, in turn, helps in stabilizing the interference. This is a very important feature since UL interference is known to be very volatile and dynamic and this result shows that radio resource management (RRM) and self-organizing network (SON) operation in general can be facilitated using DUDe.

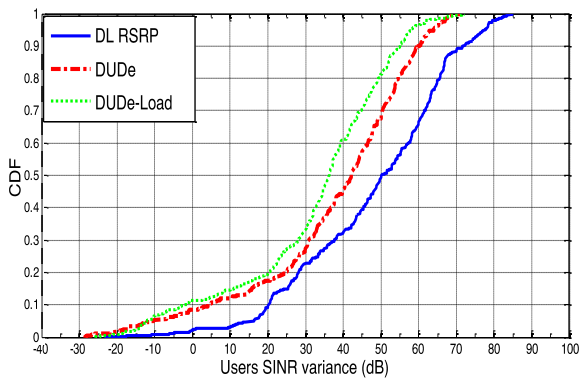

Figure 4. CDF of the SINR variance where DUDe clearly shows interference calming properties.

In *Figure 5*, we show throughput results for the interference aware power control in (4). The aim here is to try to find a trade-off between 5th and 50th percentile performance. We see, indeed, that using this power control setup we achieve a similar or even higher 5th percentile throughput as in *Setting 2* in Figure 3 where DUDe-Load outperforms DUDe by 15%. Also, in the 50th percentile the performance is similar to *Setting 1* in Figure 3 where DUDe-Load outperforms DUDe by 20%. The better performance of DUDe-Load in the 5th and 50th percentile throughputs results from the fact that the interference aware power control affects more the UEs that cause higher interference, mostly cell edge UEs, to neighboring cells while allowing the other UEs, 50th and 90th percentile UEs, to transmit with a higher power. This results in a lower interfernce scenario which benefits the cell edge UEs that are interference limited and also allows the higher achieving UEs to transmit with a higher power and, in turn, exploit the extra resources resulting from load balancing.

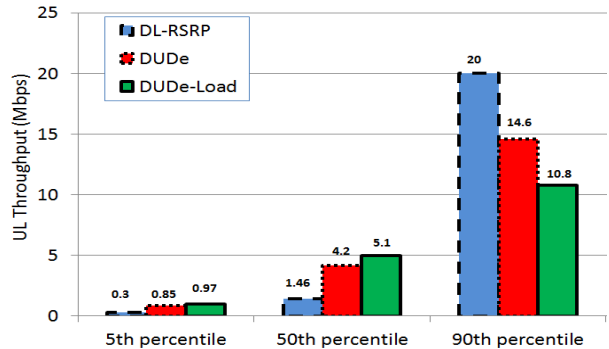

Figure 5. 5th, 50th and 90th percentile throughput of the three cases with interference aware power control according to (4).

In the results in *Figure 6* we study the throughput behaviour in the 3 cases while changing the backhaul capacity of Scells from 1 to 100 Mbps. The Mcells backhaul capacity is assumed to be 100 Mbps in all cases. We present the results for the interference aware power control setup used in *Figure 5*.

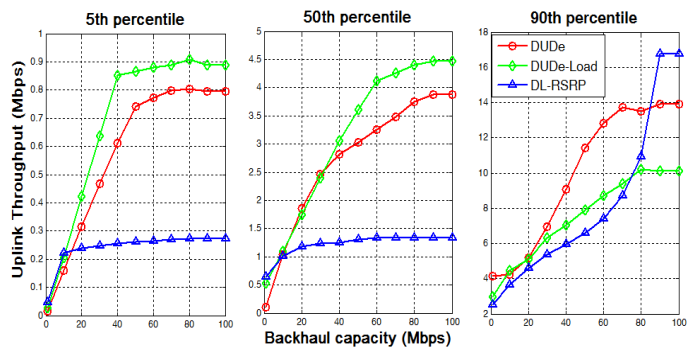

Figure 6. Throughput percentiles against backhaul capacity.

In the 5th percentile result the DUDe-load case shows the highest throughput since the UEs know of the backhaul and load capabilities of the cells. The DL-RSRP case performs better than the DUDe case up to a backhaul capacity of 10 Mbps after which DL-RSRP saturates and DUDe keeps on increasing.

Similarly, in the 50th percentile the DL-RSRP case is performing almost the same as DUDe-Load for very low Scell backhaul capacities since in the former case the UEs are mostly connected to the Mcells but as the Scell backhaul capacity increases DL-RSRP starts saturating and DUDe-Load surpasses it. Also the DUDe-load case is outperforming DUDe for the different capacities where the gain increases as the backhaul capacity of Scells increases as with the increase of Scell capacity DUDe-Load can have more options to assign UEs to Scells in a more efficient way.

Finally for the 90th percentile throughput, DUDe outperforms both DUDe-load and DL-RSRP since it has the lowest number of UEs connected to the Mcells. These UEs can get very high throughputs, up to a certain point where DL-RSRP surpasses DUDe. The reason is that Scells in DL-RSRP serve fewer UEs than the other 2 cases. Therefore after a certain backhaul capacity Scells can provide very high data rates to these UEs. Looking at the DUDe-load case, with lower Scell backhaul capacities the UEs are pushed more towards the Mcells but still DUDe-load has less UEs connected to Mcells than DL-RSRP which explains why DUDe-load outperforms DL- RSRP at the beginning but as the Scells backhaul capacity increases the load balancing role is stronger which stops the $90^{th}$ percentile throughput of DUDe-load from increasing as explained before.

Finally, in order to have some insight on the load balancing effect of DUDe we compare the variance of the number of UEs per cell in the 3 cases. This measure gives an indication of how UEs are distributed among the cells. A high variance indicates low load balancing effect and vice-versa. The variance is 470, 83 and 21 for DL-RSRP, DUDe and DUDe-load respectively. The DUDe case shows a clear improvement of load balancing over DL-RSRP which is shown by a dramatically reduced variance which, in turn, shows that the variation in the number of UEs/cell is small. The DUDe-load case shows an even lower variance (i.e. better load balancing) than DUDe as it is not only restricted on balancing the UEs between Mcells and Scells but it also improves the load balancing among Scells which is a very important feature in future ultra-dense Scell networks.

## VI. CONCLUSIONS

The decoupling of the downlink and uplink, referred to as DUDe, is an emerging paradigm shown to improve capacity significantly for cell edge users. The underlying principles of DUDe relate to a proper and independent association of the uplink and downlink. The focus of this paper has thus been to extend the prior simple association algorithms, based on the link quality in the respective links only, to a more advanced approach which considers the load in the cells as well as any backhauling constrains.

Having first introduced the general system architecture, the association algorithms as well as the simulation framework, we then presented and discussed an ample amount of performance results. The findings confirm that the enhanced DUDe achieves a reduced UL SINR variance over baseline LTE, in the order of 10-15 dB, which facilitates RRM and SON operations. Results for our load-aware DUDe show that the system throughput improves even further compared to the prior introduced baseline DUDe approach. The performance improvement depends very much on the power control mechanism used.

We have shown performance results for different power control settings and we used an interference aware power control algorithm where throughput gains of the load aware DUDe over baseline DUDe are 15% and 20% in the $5^{th}$ and $50^{th}$ percentile throughput respectively.

We believe that the DUDe technique is a strong candidate for 5G architecture designs and it can be very useful in many applications like real-time video gaming, Machine Type Communications (MTC), among others, where uplink optimization is very critical. Our future work will focus on specific scheduling algorithms tailored to DUDe which are able to maximize capacity of the overall system in both uplink and downlink whilst considering the constrained backhaul.

ACKNOWLEDGMENT

This research has been co-funded by Vodafone Group R&D and CROSSFIRE (un-CooRdinated netwOrk StrategieS for enhanced interFerence, mobIlity, radio Resource, and Energy saving management in LTE-Advanced networks) MITN Marie Curie project.